\documentclass[aps,prd,preprint,nofootinbib,a4paper,11pt,superscriptaddress]{revtex4-1}

\usepackage[T1]{fontenc}
\usepackage[centertags]{amsmath}
\usepackage{amsfonts}
\usepackage{amssymb}
\usepackage[sort&compress]{natbib}
\usepackage[hypertex]{hyperref}
\usepackage[dvips]{graphicx}

\newcommand{\e}{\varepsilon}

\newcommand{\al}{\alpha}

\newcommand{\la}{\lambda}
\newcommand{\La}{\Lambda}

\begin{document}

\title{Comment on ``Quantum versus classical instability of scalar fields in curved backgrounds'' [arXiv:1310.2185]}

\date{\today}

\author{P.O. Kazinski}
\email[E-mail:]{kpo@phys.tsu.ru}
\affiliation{Physics Faculty, Tomsk State University, Tomsk, 634050 Russia}
\affiliation{Laboratory of Mathematical Physics, Tomsk Polytechnic University, Tomsk, 634050 Russia}

\begin{abstract}

I show that the claim of the paper [arXiv:1310.2185] on the absence of instability for a minimally coupled scalar field on a static spherically symmetric gravitational background is  incorrect.

\end{abstract}


\maketitle

The authors of the paper \cite{MMV} claim that ``... there are no tachyonic modes for minimally coupled scalar fields in asymptotically flat spherically symmetric static spacetimes containing no horizons...'' (p. 2 and Appendix of \cite{MMV}). In fact, the authors tried to prove that there is no any Jeans instability in this case contrary to the results of \cite{psfss}. This statement is, of course, incorrect as I shall show below.

Consider a static spherically symmetric metric [Eq. (1), \cite{MMV}]
\begin{equation}
    ds^2=-e^{2\Xi(r)}dt^2+e^{2\La(r)}dr^2+r^2(d\theta^2+\sin^2\theta d\phi^2).
\end{equation}
Changing the variable [Eq. (14), \cite{MMV}]
\begin{equation}
    x(r)=\int_0^r dr' e^{\La(r')-\Xi(r')},
\end{equation}
the authors reduce this interval to
\begin{equation}
    ds^2=e^{2\Xi}(-dt^2+dx^2)+r^2d\Omega^2.
\end{equation}
Then, the variables in the massless KG equation,
\begin{equation}
    (\nabla^2-\xi R)\Phi=0,
\end{equation}
are separated,
\begin{equation}
    \Phi=e^{-i\omega t}\frac{\psi_{\omega l}(r)}{r}Y_{lm}(\theta,\phi),
\end{equation}
and the radial part of the KG equation becomes
\begin{equation}
    -\psi''+\Big[\frac{r''}{r} +e^{2\Xi}\big(\frac{l(l+1)}{r^2}+\xi R\big)\Big]\psi=\omega^2\psi,
\end{equation}
where all the derivatives are taken with respect to $x$. For the $s$-wave ($l=0$) and for the minimal coupling, $\xi=0$, it becomes
\begin{equation}\label{KG_smode}
    -\psi''+\frac{r''}{r}\psi=\omega^2\psi,
\end{equation}
As follows from Eq. (14), the function $r(x)$ is an arbitrary smooth function on the interval $x\in[0,+\infty)$, but obeying the restrictions
\begin{enumerate}
  \item $r'(x)>0$,\;\;$x\in[0,+\infty)$,
  \item $\lim_{x\rightarrow+\infty} r'(x)=1$.
\end{enumerate}
In Appendix of \cite{MMV}, the authors provide an obscure reasoning that Eq. \eqref{KG_smode} has no bound states for any $r(x)$ satisfying the requirements above. If the latter took place then the general statement proposed by the authors (see the first paragraph) would hold.

However, it is clear from \eqref{KG_smode} that one can simply draw the monotonic function $r(x)$ satisfying all the requirements and possessing a negative second derivative on a sufficiently large interval such that the effective potential,
\begin{equation}
    V_{eff}=r''/r,
\end{equation}
admits bound states. To be more specific, I shall give the explicit example.

\begin{figure}[t]
\centering
\includegraphics*[width=8cm]{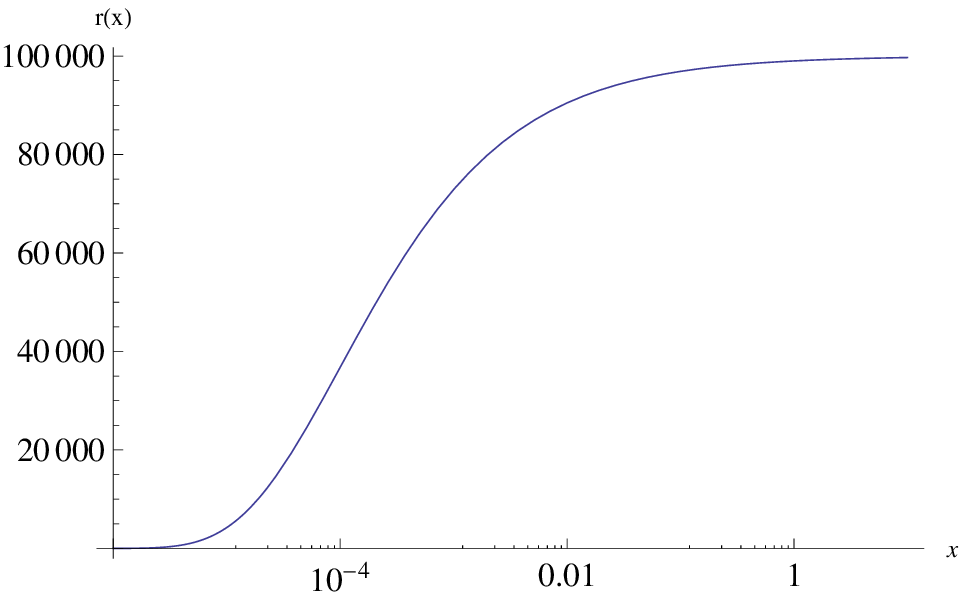}\;
\includegraphics*[width=8cm]{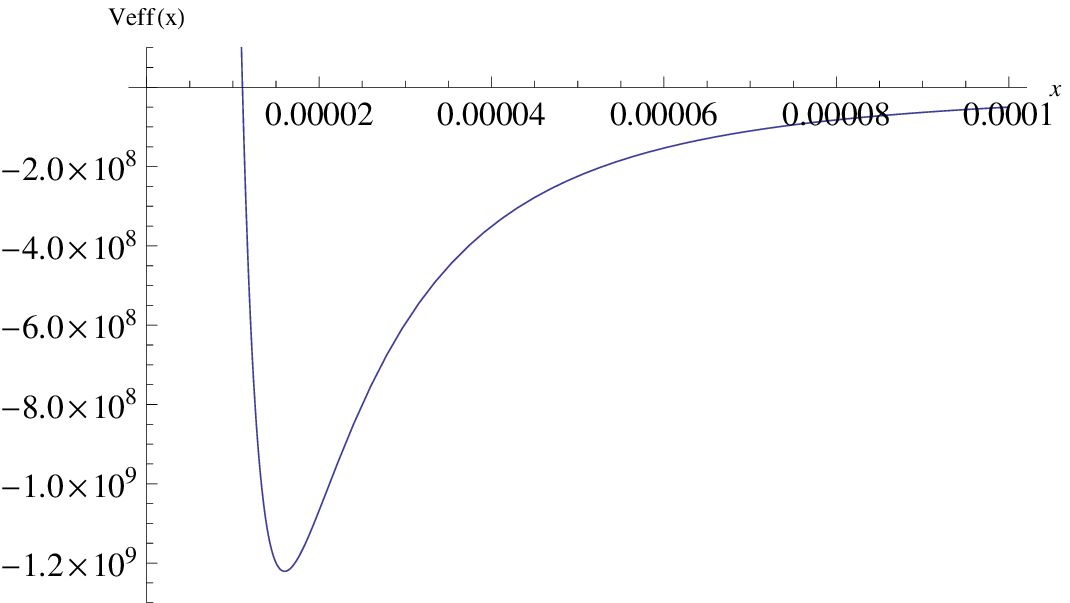}
\caption{Left panel: The plot of $r(x)$ with  $a=10^{-2}$, $b=10^5$, and $\al=1/2$. Right panel: The plot of $V_{eff}(x)=r''/r$ with $a=10^{-2}$, $b=10^5$, and $\al=1/2$.}
\label{plot}
\end{figure}

Consider the function
\begin{equation}
    r(x)=x+b e^{-a/x^\al},\quad a,b,\al>0.
\end{equation}
It complies with all the requirements above. I take $a=10^{-2}$, $b=10^5$, and $\al=1/2$. The plots of $r(x)$ and $V_{eff}$ with these parameters are given in Fig. \ref{plot}. The profile of the effective potential resembles the Lennard-Jones one. In order to prove that $V_{eff}$ possesses the bound states, I use the quasiclassical criterion (the Borh-Sommerfeld quantization rule)
\begin{equation}\label{qc_criter}
    \int_{x_1}^{x_2}dx\sqrt{-r''/r}\geq\pi/2,
\end{equation}
where $x_1$ and $x_2$ are the turning points. If the turning point $x_1=0$ then the RHS of this inequality should be replaced by $\pi/4$. In our case, one can find that $x_1\approx1.11\times10^{-5}$, $x_2=+\infty$, and the numerical evaluation of the integral \eqref{qc_criter} results in
\begin{equation}
    \int_{x_1}^{+\infty}dx\sqrt{-r''/r}\approx 4.7>\pi/2.
\end{equation}
Thus the bound state exists. Curiously enough, this example can be treated analytically at very large $b$. In this case, in evaluating the integral \eqref{qc_criter}, one may replace
\begin{equation}
    r(x)\approx be^{-a/x^\al}\;\Rightarrow\;V_{eff}=\frac{\al^2 a^2}{x^{2\al+2}}-\frac{\al(\al+1)a}{x^{\al+2}},\qquad x_1=\Big(\frac{\al a}{\al+1}\Big)^{1/\al}.
\end{equation}
The LHS of \eqref{qc_criter} equals $\pi(1+\al^{-1})/2$ and can be made as large as one wants by choosing sufficiently small $\al$. In particular, in the case $\al=1/2$ we have $3\pi/2\approx4.7$. Of course, one may object that the example above is somewhat unphysical. Nevertheless it provides a counterexample to the general statement of \cite{MMV}.

Moreover, one can prove that the instability (tachyonic modes) exists even in a more physically reasonable situation. Using the fact that
\begin{equation}
    r''/r=e^{2(\Xi-\La)}\frac{\Xi'_r-\La'_r}{r}=r'^2\frac{\Xi'_r-\La'_r}{r},
\end{equation}
the integral \eqref{qc_criter} can be rewritten as
\begin{equation}
    \int_{x_1}^{x_2}dx\sqrt{-r''/r-e^{2\Xi}\xi R}=\int_{r_1}^{r_2}dr\sqrt{(\La'_r-\Xi'_r)/r-e^{2\La}\xi R},
\end{equation}
where the nonminial coupling has been restored. Then, employing the Einstein equations for a static spherically symmetric metric \cite{LandLifshCTF}, we have
\begin{equation}
    \int_{x_1}^{x_2}dx\sqrt{-r''/r-e^{2\Xi}\xi R}=\int_{r_1}^{r_2}dr\Big[-4\pi Ge^{2\La}(T^t_t+T^r_r-2\xi T)+(1-e^{2\La})/r^2 \Big]^{1/2}\geq\pi/2.
\end{equation}
At large $r$, one can neglect the terms at $r^{-2}$ and obtain
\begin{equation}\label{instab}
    \int_{r_1}^{r_2}dr e^\La\Big[-4\pi G(T^t_t+T^r_r-2\xi T)\Big]^{1/2}\geq\pi/2.
\end{equation}
Introducing the Jeans length (see, e.g., \cite{GorbRub}),
\begin{equation}
    \la_J:=\Big(\frac{\pi c_s^2}{G\e}\Big)^{1/2},
\end{equation}
where $\e$ is the energy density and $c_s$ is the sound speed in the matter, which I assume to be a fluid with
\begin{equation}
    p=c_s^2\e,
\end{equation}
one can estimate from \eqref{instab} the characteristic wavelengths $\ell$ of unstable modes. Roughly,
\begin{equation}\label{ell}
    \ell\geq\frac{\la_J}{4c_s\sqrt{1-2\xi-(1-6\xi)c_s^2}},
\end{equation}
in the weak field limit. The case $\xi=1/6$ corresponds to the conformal coupling. If the radicand in \eqref{ell} is nonpositive then one may expect that there is no such an instability. However, we see that for an ultrarelativistic fluid, $c_s^2=1/3$, this instability takes place for any $\xi$. It also always exist at $\xi\in[0, 1/2]$ and, in particular, for the minimal and conformal couplings. Thus the Bose-Einstein condensation discussed in \cite{psfss} exist even in the weak field limit for the minimal coupling contrary to the  claims of \cite{MMV}. Such a condensation develops on the scales comparable with the structure formation scale of the Universe.

\end{document}